\documentclass[epj,twocolumn]{webofc}
\usepackage[varg]{txfonts}   % Web of Conferences font
\usepackage{subfigure,caption}
\woctitle{Hadron Collider Physics symposium 2012}
\begin{document}
%\preprint{Cavendish-HEP-2013/02}
\title{Polarisation of electroweak gauge bosons at the LHC}
%\hfill\hfill \text{Cavendish-HEP-2013/02}

\author{James Stirling\inst{1} \and Eleni Vryonidou\inst{1}\fnsep\thanks{\email{hv222@hep.phy.cam.ac.uk}} 
\hfill\hfill \text{Cavendish-HEP-2013/02}}
        % etc.

\institute{Cavendish Laboratory, J.J. Thomson Avenue, Cambridge CB3 0HE, UK}

\abstract{We present results for the polarisation of gauge bosons produced at the LHC. Polarisation
effects for $W$ bosons manifest themselves in the angular distributions of the lepton and
in the distributions of lepton transverse momentum and missing transverse energy. The polarisation 
is discussed for a range of different processes producing
$W$ bosons such as $W+$jets and $W$ from top production. The relative contributions of the
different polarisation states vary from process to process, reflecting the dynamics of the
underlying hard-scattering process. We also calculate the polarisation of the $Z$
boson produced in association with QCD jets at the LHC.
}
\maketitle
\section{Introduction}
\label{intro}
The production of vector bosons has been extensively studied at past and present high-energy collider experiments as it is 
important both for confirming Standard Model (SM) electroweak predictions and in the
search for evidence of New Physics. Recent LHC analyses have studied many different processes involving $W$ and $Z$ bosons. 

On the theory side $W, Z +$ jets production has been extensively
studied in the literature, with NLO corrections recently calculated for up to four associated
jets \cite{Berger:2010zx,Ita:2011wn}. Another important tool in
the search for New Physics is the use of underlying properties to distinguish potential signals from the SM background. In particular, the use of cross-section ratios is beneficial as these
 suffer less from theoretical and experimental uncertainties. One such property is the polarisation of gauge bosons and
the resulting distributions of lepton transverse momentum and missing transverse energy.

Measurements of the polarisation of $W$ bosons have been
undertaken by previous high-energy collider experiments. The polarisation of $W$ bosons from top
pair production and decay has been measured by the
Tevatron experiments \cite{Abazov:2010jn,Aaltonen:2010ha} and the results have been employed to set limits on anomalous $Wtb$ couplings ~\cite{AguilarSaavedra:2006fy}.
The polarisation of weak bosons produced in pairs has been extracted by LEP experiments in \cite{Achard:2002bv,Abbiendi:2003wv} and used to set limits on anomalous triple gauge boson
couplings in \cite{Abdallah:2008sf}. The angular distributions of the produced
leptons and the corresponding angular coefficients have also been investigated experimentally
 and phenomenologically, see for example \cite{Collins:1977iv,Lam:1980uc,Mirkes:1992hu,Mirkes:1994eb}.

At the LHC, new kinematical regions are accessible,
polarisation can be used as a probe of the underlying
interactions in regions never explored before.
$W$ bosons produced along the beam direction have long been known to be predominantly
left-handed at the LHC \cite{Ellis:1991qj}. In this case the polarisation
of the $W$ is simply determined by the momentum of the colliding quark and antiquark.
The $W$ boson is left-handed if the quark has more momentum than the antiquark, and
right-handed otherwise. The exact values of the fractions at zero transverse momentum therefore
depend on the relative values of the corresponding quark and antiquark PDFs. At the LHC,
quarks have on average more momentum than antiquarks which explains the difference in
values $f_L =$ 0.73 and $f_R =$ 0.27 for $W^+$, while the same arguments apply for $W^-$ leading
to $f_L = $0.68 and $f_R =$ 0.32. At the Tevatron, quarks and antiquarks have on average
the same momentum and therefore we expect the fractions to be closer to 0.5 for $W^+$, calculated 
indeed to be  $f_L =$ 0.60 and $f_R =$0.40. The fractions for $W^-$ are
exactly reversed, with $W^-$ being preferentially right-handed.
In \cite{Bern:2011ie} it was observed that $W$ bosons produced
at the LHC in association with QCD jets are in general preferentially left-handed.
 The dependence of the polarisation on the number of jets and the NLO
pQCD corrections has also been investigated, with the results shown to be rather stable.
The CMS \cite{Chatrchyan:2011ig} and ATLAS \cite{Aad:2012nn} collaborations at the LHC have measured the polarisation of $W$ bosons
produced in association with QCD jets at large transverse momentum, and demonstrated
good agreement with the SM predictions first presented in \cite{Bern:2011ie}. 
\vspace{-0.2cm}
\section{$W$ polarisation in $W+$jets}
\label{Wjets}The polarisation changes the lepton and missing
energy distributions in the leptonic $W$ decays and it was this observation that led the authors of \cite{Bern:2011ie} 
to conclude that the polarisation of $W$ bosons at large
transverse momentum is predominantly left-landed. The characteristic shape of the ratio of the charged lepton $p_T$ distribution to
the missing $E_T$ distribution is shown in Fig. 1 for the LHC at 7 TeV. In this plot
 the predominantly left-handed polarisation translates into a decreasing
ratio for $W^+$ and an increasing ratio for $W^-$.
\begin{figure}
\centering
\includegraphics[scale=0.53,clip]{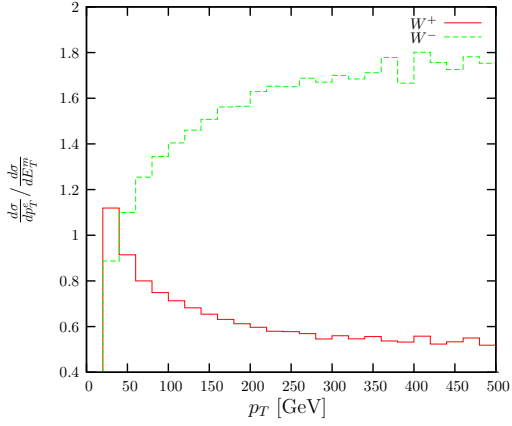}
\caption{Ratio of the differential distribution of the lepton transverse momentum to the distribution of the 
missing transverse energy for both $W^+$ and $W^-$ at 7~TeV.}
\label{fig-1}
\vspace{-0.4cm}
\end{figure}

To obtain a quantitative measure of the polarisation we use the angular distribution of the $W^+$ decay products in the $W^+$ rest 
frame described by:
\begin{equation}
 \frac{1}{\sigma}\frac{d\sigma}{d\text{cos}\theta^*}=\frac{3}{8}(1-\text{cos}\theta^*)^2 f_L+\frac{3}{8}(1+\text{cos}\theta^*)^2f_R+\frac{3}{4}\text{sin}^2\theta^*f_0,
\label{diffeq}
\end{equation}where $\theta^*$ is the angle in the $W$ rest frame between the charged lepton and the $W$ flight direction in the lab frame, and $f_{0,L,R}$ are the polarisation fractions. For $W^-$, $f_R$ and $f_L$ are interchanged.
The normalisation is chosen so that $f_0+f_L+f_R=1$ and any dependence on the azimuthal angle has been integrated out.
In \cite{Bern:2011ie} it is noted that $\sigma$ in Eq.~(\ref{diffeq}) can be any differential cross section that does not depend 
on the kinematics of the individual leptons. 

Based on Eq.~(\ref{diffeq}), in order to obtain the polarisation fractions in the helicity 
frame we use the following expressions (for $W^+$): 
\begin{eqnarray}
f_0&=&2-5\langle \rm{cos}^2\theta^{*}\rangle,\label{f0}\\
f_L&=&-\frac{1}{2}-\langle \rm{cos}\theta^*\rangle+\frac{5}{2}\langle \rm{cos}^2\theta^{*}\rangle, \label{fL}\\
f_R&=&-\frac{1}{2}+\langle \rm{cos}\theta^*\rangle+\frac{5}{2}\langle \rm{cos}^2\theta^{*}\rangle.
\label{fR}
\end{eqnarray} 
These functions of $\theta^*$ can be used on an event-by-event basis as projections to extract the 
polarisation fractions when full acceptance for the leptons applies. 

\begin{figure}[h]
\centering
\subfigure[]{
\includegraphics[trim=2.4cm 0 0 0,scale=0.53]{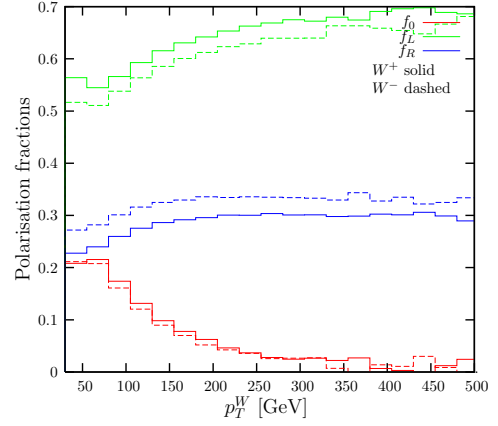}}
\label{fig7a}
\centering
\subfigure[]{
\includegraphics[trim=1.4cm 0 0 0,scale=0.53]{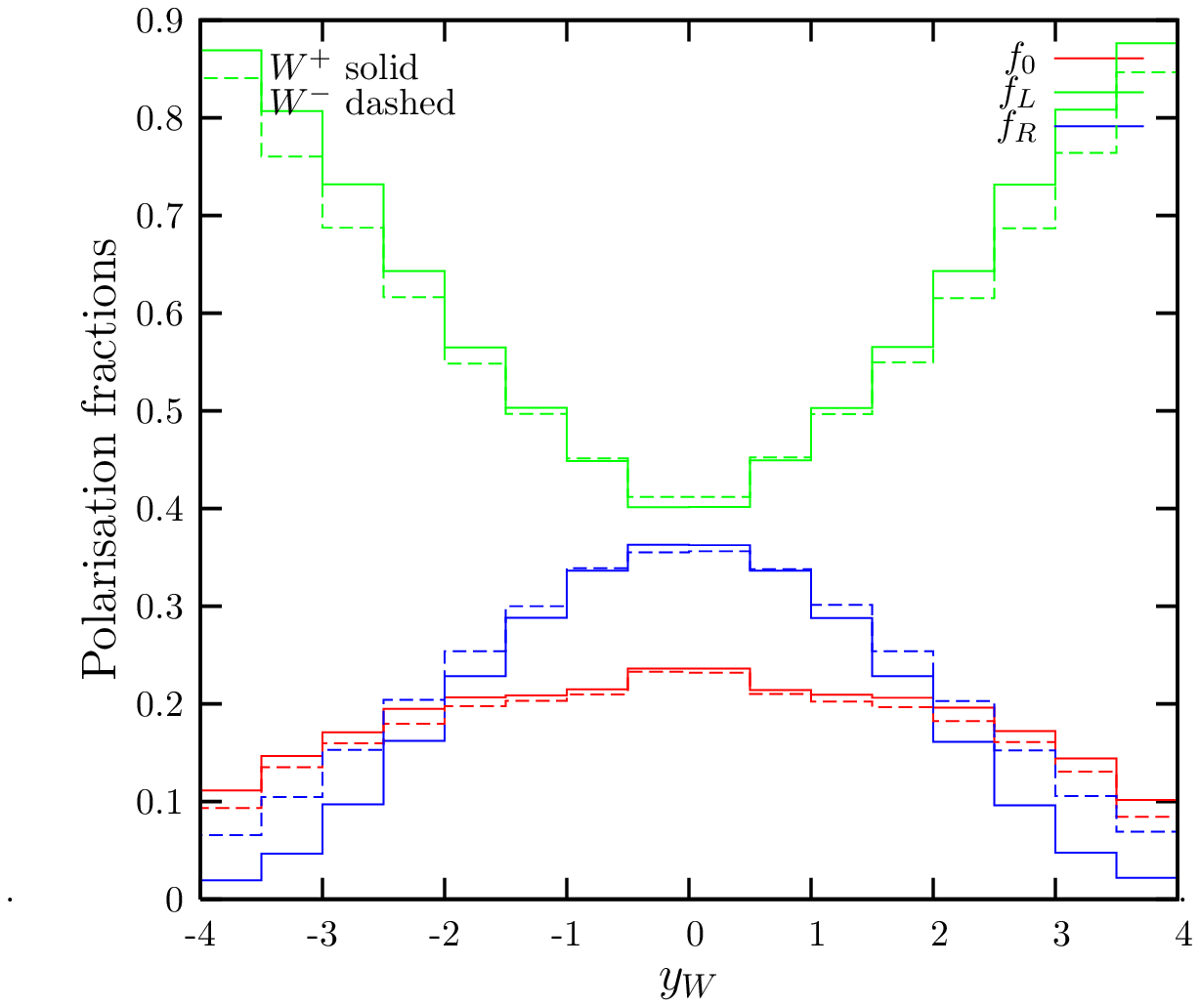}
}
\caption{Polarisation fractions as a function of (a) $p_T^W$ and (b) $y_W$ for 7 TeV and a jet
$p_T$ (=$p_T^W$ for $W+1$ jet) cut of 30 GeV for both $W^+$ and $W^-$.}
\label{fig7b}
\vspace{-0.4cm}
\end{figure}

In \cite{Stirling:2012zt} we obtained the polarisation fractions for $W+1$ jet with a simple jet
$p_T$ cut of 30~GeV for the LHC at 7~TeV for both $W^+$ and $W^-$, shown in Fig.~\ref{fig7b} as a function of the $W$ transverse momentum and rapidity. The small difference between the $W^+$ and $W^-$ polarisations
is due to the difference between the valence $u-$ and $d-$quark PDFs that
forces the $W^+$ to be slightly more left-handed. We also see the longitudinal fraction falling to zero at large transverse momentum, in agreement with the equivalence theorem. 

\begin{figure}[h]
 \centering
 \includegraphics[trim=1.7cm 0 0 0,scale=0.53]{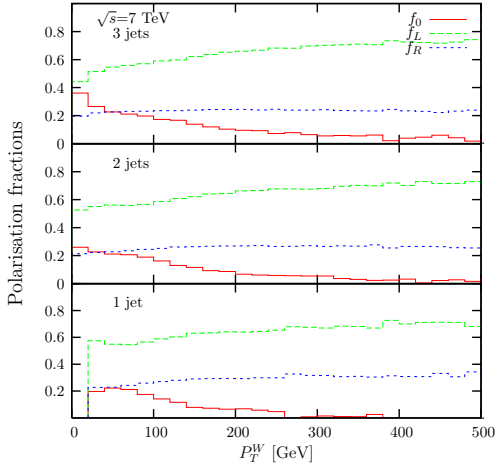}
 \caption{Polarisation fractions for $W^+$+1, 2 and 3~jets with $p_T^j>30$~GeV at 7~TeV obtained using MCFM.}
\label{morejets}
 \end{figure}

For comparison we show in Fig.~\ref{morejets} the results of the polarisation fractions for 1, 2 and 3 jets with a jet transverse momentum 
cut of 30~GeV obtained using MCFM \cite{MCFM}. As already noted in
\cite{Bern:2011ie}, even though the kinematics quickly
 become more complicated with an increasing number of jets the polarisation fractions are not very sensitive to the number
of jets, with changes only observed at low $W$ transverse momentum and the polarisation remaining predominantly left-handed 
at large transverse momentum.

In addition to the number of associated jets, the dependence of the polarisation fractions on higher-order effects can be investigated. NLO corrections are found to have no sizable impact on the polarisation fractions. The definition of the polarisation 
fractions as effectively ratios over the total cross sections helps reduce the sensitivity to the NLO corrections. 
In \cite{Bern:2011ie} it has been shown using SHERPA that the results remain stable 
even when parton shower effects are taken into account.

\begin{figure}[h]
\subfigure[]{
 \includegraphics[trim=1.3cm 0 0 0,scale=0.51]{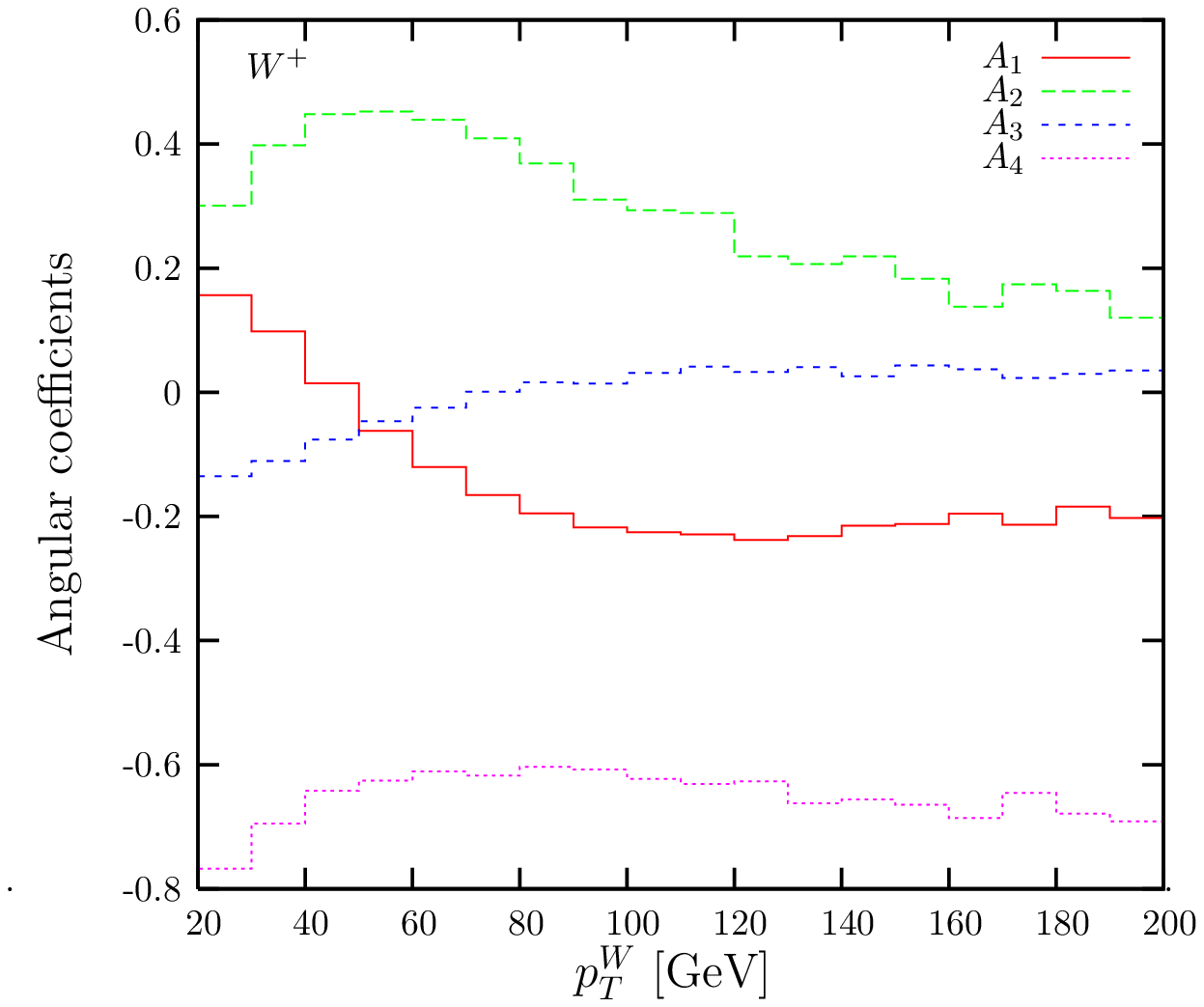}}
\subfigure[]{
 \includegraphics[trim=1.3cm 0 0 0,scale=0.51]{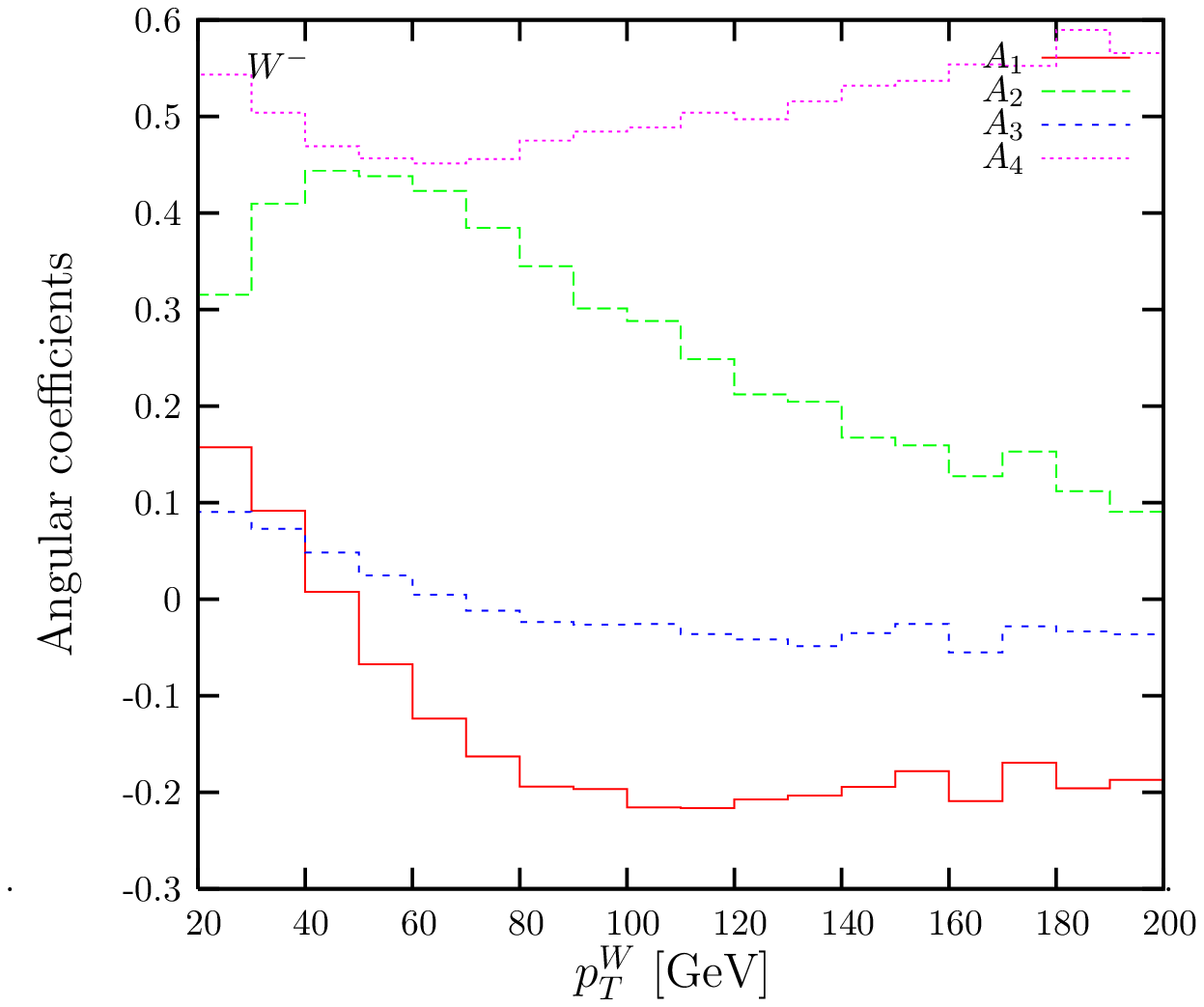}}
 \caption{Angular coefficients  for $W^-$ and $W^+$ plus one jet with no imposed cuts 
in proton-proton collisions at 7 TeV. }
 \label{ang}
 \end{figure}

The polarisation fractions constitute the diagonal elements of the $W$ spin-density matrix. One can study the off diagonal elements by introducing the azimuthal angle. The differential cross section is then written as: 
\begin{eqnarray}\nonumber
\hspace*{-1cm} \frac{1}{\sigma}\frac{d\sigma}{d\text{cos}\theta^*d\phi^*}&=&\frac{3}{16\pi}[(1+\text{cos}^2\theta^*)+A_4\text{cos}\theta^*\\\nonumber
&+&A_0\frac{1}{2}(1-3\text{cos}^2\theta^*)+A_1 \text{sin} 2\theta^* \text{cos}\phi^*\\
&+&A_2\frac{1}{2}\text{sin}^2\theta^*\text{cos}2\phi^*+A_3\text{sin}\theta^*\text{cos}\phi^*],
\label{fulldist}
\end{eqnarray} with the angle $\phi^*$ defined as in \cite{Bern:2011ie}. At LO five coefficients are non-zero. At NLO two more appear but these remain close to zero. One can relate the polarisation fractions $f_{L,R,0}$ to $A_0$ and $A_4$ by integrating Eq.~(\ref{fulldist}) over $\phi^*$. We show the LO results obtained for $W+1$~jet for $W^-$ and $W^+$ in Fig.~\ref{ang}.

The first measurement of the
polarisation of the $W$ boson by CMS~\cite{Chatrchyan:2011ig}
 introduced the variable $L_p$, defined as: 
\begin{equation}
L_p=\frac{\vec{p}_T(l)\cdot \vec{p}_T(W)}{|\vec{p}_T(W)|^2},
\end{equation} 
where all quantities can in principle be reconstructed. In the limit of very high $W$ transverse momentum 
$\cos\theta^*=2(L_p-1/2)$. The results in \cite{Chatrchyan:2011ig} are obtained using the template 
method to extract the distributions of $L_p$ for pure longitudinal, left- and right-handed samples and then fitted to the data to obtain the polarisation fractions. This accounts for the effect of the selection cuts. Similarly ATLAS introduces the angle:
\begin{equation}
\textrm{cos}\theta_{2D}=\frac{\vec{p}^*_T(l)\cdot \vec{p}_T(W)}{|\vec{p}^*_T(l)||\vec{p}_T(W)|},
\end{equation}which is a 2D projection of the angle $\theta^*$ and the analysis uses the template method to extract the polarisation results. Both CMS and ATLAS have presented results for the 2010 data set. The results show reasonable agreement with the theoretical predictions given the uncertainties.
\section{$Z$ polarisation in $Z+$jets}
\begin{figure}[h]
\centering
\includegraphics[trim=1.2cm 0 0 0,scale=0.51]{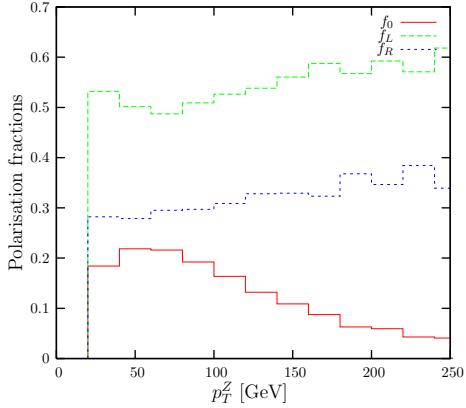}
\caption{$Z$ polarisation fractions from the electron decay channel at 7 TeV for $Z+1$~jet.}
\label{nuel}
%\end{minipage}
\end{figure}

The methods used for $W$ bosons can in principle also be used to measure the polarisation of 
$Z$ bosons produced at the LHC.
For the $Z$ bosons decaying to two charged leptons, the shape of the positively and negatively charged 
lepton distributions cannot be used as a direct probe of polarisation in the same way as for the $W$ boson, because the $Z$ 
couples to {\it both} right- and left-handed fermions. 
Therefore the SM leptonic couplings need to be taken into account and the equivalent of Eq.~(\ref{diffeq}) for $Z$ decay 
to a pair of fermions is:
\begin{eqnarray}\nonumber
 \frac{1}{\sigma}\frac{d\sigma}{d\text{cos}\theta^*}&=&\frac{3}{8}\bigg(1+\text{cos}^2{\theta^*}-
\frac{2(c_L^2-c_R^2)}{(c_L^2+c_R^2)}\text{cos}\theta^*\bigg) f_L\\\nonumber
&+&\frac{3}{8}\bigg(1+\text{cos}^2{\theta^*}
+\frac{2(c_L^2-c_R^2)}{(c_L^2+c_R^2)}\text{cos}\theta^*\bigg)f_R\\
&+&\frac{3}{4}\text{sin}^2\theta^*f_0,
\label{diffeqZ}
\end{eqnarray}
 with $c_R$ and $c_L$ the right- and left-handed couplings of the fermion to the $Z$ and 
$\theta^*$ the angle measured in the $Z$ rest frame between the antiparticle
 and the $Z$ flight direction in the lab frame. Based on Eq.~(\ref{diffeqZ}) we can use appropriate projections to obtain  the polarisation fractions:
\begin{eqnarray}
f_0&=&2-5\langle \rm{cos}^2\theta^{*}\rangle,\\
f_L&=&-\frac{1}{2}-\frac{(c_L^2+c_R^2)}{(c_L^2-c_R^2)}\langle \rm{cos}\theta^*\rangle+\frac{5}{2}\langle \rm{cos}^2\theta^{*}\rangle, \\
f_R&=&-\frac{1}{2}+\frac{(c_L^2+c_R^2)}{(c_L^2-c_R^2)}\langle \rm{cos}\theta^*\rangle+\frac{5}{2}\langle \rm{cos}^2\theta^{*}\rangle.
\end{eqnarray} 
For the decay to neutrinos the projections are identical to those for $W$ bosons, since $c_R^{\nu}=0$. 
In contrast to the case of $W$ bosons, for which $\theta^*$ cannot be extracted precisely due to the unreconstructed 
longitudinal $W$ momentum, for $Z$ production $\theta^*$ is unambiguously defined and reconstructed. 
Therefore there is no need to use the variables $L_p$ and $\theta_{2D}$ introduced earlier. 
The polarisation results are shown in Fig.~\ref{nuel} for $Z+1$~jet at 7 TeV with a jet $p_T$ cut of 30~GeV. 
Evidently $Z$ bosons are also predominantly left-handed at non-zero transverse momentum. 
Comparing the results for $Z$ to those for $W+1$~jet  we note that while the $Z$ is also predominantly left-handed, 
the exact values of the fractions differ, as they arise from a combination of the different quark flavour combinations 
producing the $Z$ boson and the fact that the $Z$ couples to both left- and right-handed quarks. The analysing power for the $Z$ polarisation is only $(c_L^2-c_R^2)/(c_L^2+c_R^2)=$15\%, which makes distinguishing left and right 
polarisation more difficult than the $W$ case where the analysing power is 100\%. 
A complication that arises in $Z$ boson studies is the small admixture of photon events. 
In terms of the expressions given above, these apply only to pure $Z$ exchange. 
Experimentally this is limited by introducing a cut constraining the lepton pair invariant mass to be close to the $Z$ mass.

\section{Polarisation of $W$ in other processes}
In addition to the production of $W$ bosons in association with QCD jets,
 the polarisation properties of $W$ bosons from other sources can be investigated. 
The polarisation of $W$ bosons from top pair production has been studied in the literature. 
The projections used differ from those used for $W+$~jets, 
as the angle $\theta^*$ is defined in the $W$ rest frame relative to the $W$ direction in the top rest frame.

Using this definition we find that for $W^+$ bosons: $f_0=0.70$ and $f_L=0.30$. These can be extracted from analytic expressions 
involving the $W$ and top masses (for massless $b-$quarks):
\begin{equation}
f_0=\frac{m_t^2}{m^2_t+2m_W^2} \,\,\,\,\, \textrm{and} \,\,\,\, f_L=\frac{2m_W^2}{m^2_t+2m_W^2}.
\end{equation}

In this case we find that the polarisation fractions have
 no dependence on the $W$ $p_T$, which follows from the way polarisation is defined using the top rest frame. 

In this process $W^+$ and $W^-$ are exactly equivalent as these are always produced in pairs from the decaying top--anti-top 
pair, hence the polarisation
fractions are related by $f_0^+=f_0^-$ and $f_R^+=f_L^-$. Due to the symmetric production mechanism we have: 
\begin{equation}
\frac{{d\sigma(t\bar{t})}/{dp_T^{e^+}}}{{d\sigma(t\bar{t})}/{dp_T^m}}
=\frac{{d\sigma(t\bar{t})}/{dp^{e^-}_T}}{{d\sigma(t\bar{t})}/{dp_T^m}}.
\end{equation}
Also, no asymmetry is expected
between the number of produced electrons and positrons. 

\begin{figure}[h]
 \centering
 \includegraphics[trim=1.3cm 0 0 0,scale=0.57]{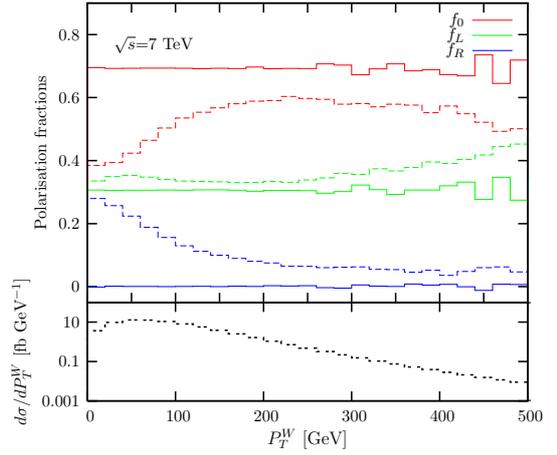}
 \caption{Polarisation fractions for $W^+$ from top decays with no imposed cuts
using the two angle definitions. Solid: top rest frame and dashed: lab frame.}
 \label{tnew}
 \end{figure}
To be consistent with our analysis for  $W+$~jets we should of course use the same projections
with $\theta^*$ defined relative to the $W$ direction in the lab frame. The comparison
between the two definitions is shown in Fig.~\ref{tnew}, together with the distribution
for the $W$ $p_T$. For $W^-$ $f_R \Leftrightarrow  f_L$. It is clear from the plot that the polarisation fractions are highly sensitive to the definition of the angle and therefore frame dependent. Use of the lab frame in the definition also introduces a dependence of the fractions on the $p_T$ of the $W$ boson.

In addition to $W$ from top pair production, in \cite{Stirling:2012zt} we present results for other processes at the LHC such as $W+Z$ and $W+H$ associated production, $W$ pair production, $W$ from single top decay using the same projections as 
for $W+$~jets. We select the available MCFM subprocesses where 
the $W^+$ decays to a positron and a neutrino and the other particles produced in the event ($W^-$, $H$, $Z$) decay 
hadronically.

We find that the polarisation fractions of the produced $W^+$ boson vary from process.
These processes are in general a subdominant source of $W$ bosons compared to QCD $W +$~jets but it is still 
 important to explore them both because they constitute a further set of backgrounds for New Physics searches 
and also because they are interesting  processes in their own right. The cross-section results for a range of processes at 
7~TeV and the total polarisation fractions are given in Table~\ref{smtable}.  The only cut imposed for the 
results shown in Table \ref{smtable} is a jet $p_T$ cut of 30~GeV for the $W+$~jets processes. 
As the cross sections for the non-QCD processes are much smaller, the experimental determination of 
the polarisation fractions could well be impeded by low statistics.
\begin{table} [h]
\caption{Comparison of results for the different processes leading to $W^+$ production. }
\begin{center}
    \begin{tabular}{ | c | c | c | c | c |}
    \hline
   Process  & $\sigma$ [fb] & $f_0$ & $f_L$ & $f_R$   \\ \hline
    $W+1$ jet($p^j_T>30$ GeV) & 6.11$\cdot 10^5$ & 0.20 & 0.56 & 0.24 \\ \hline
    $W+2$ jets($p^j_T>30$ GeV) & 2.15$\cdot 10^5$& 0.20 & 0.56 & 0.23\\ \hline
    $W+3$ jets($p^j_T>30$ GeV) & 0.74$\cdot 10^5$  & 0.21 & 0.56 & 0.23  \\ \hline
    $t(\rightarrow b W^+)\bar{t}(\rightarrow \bar{b}q\bar{q}$) &  9320 & 0.46 & 0.37 & 0.17 \\ \hline
    Single top($t$-channel) &  4067 & 0.42 & 0.43 & 0.15  \\ \hline
    Single top($s$-channel) &  205 & 0.24 & 0.61 & 0.14  \\ \hline
    $W+Z$($Z\rightarrow q\bar{q}$)& 944 & 0.05 & 0.72 & 0.23 \\  \hline
    $W+H$($H\rightarrow b\bar{b}$) & 53 & 0.05 & 0.72 & 0.23 \\  \hline
    $W$ pair($W^-\rightarrow q\bar{q}$) & 2048 & 0.26 & 0.48 & 0.25  \\  \hline
   \end{tabular}
\end{center}
\label{smtable}
\end{table} 

Plots of the polarisation fractions as a function of the $W$ $p_T$ for these processes are shown in \cite{Stirling:2012zt}. It is clear from the plots and from Table~\ref{smtable} that the polarisation properties of $W$ bosons 
depend strongly on the production process, as they are determined by the underlying interaction.

\section{Conclusions}
We have studied the polarisation of electroweak gauge bosons at the LHC. We have seen that $W$ bosons produced in association with QCD jets at 
non-zero transverse momentum are preferentially left-landed at the LHC. This leads to asymmetries between the 
charged lepton and neutrino transverse momentum distributions. 

We have compared the polarisation of $W$ bosons produced with QCD jets to $W$ bosons from top pair production and decay,
 calculating the polarisation fractions. 
For $W$ from top pair production we have compared the polarisation fractions obtained in two different frames, to show that the polarisation fractions 
are strongly frame dependent. We remark that the $f_0=0.7$, $f_L=0.3$ fractions often mentioned in the literature 
are valid only when defining the relevant angle in the top rest frame.
  
We have used the same procedure for other $W$ producing processes at the LHC to study the polarisation of $W$ bosons 
as a function of the $W$ transverse momentum. The origin of the 
different results for different processes is related to the underlying physics of the interaction and the helicities of the other particles involved. 
Therefore a study of the polarisation properties can be used in conjunction with kinematics to distinguish between 
different sources of $W$ bosons. This is also helpful for New Physics searches where new interactions might give 
different polarisation fractions and can therefore be used as a handle to disentangle the signal from the SM background. Other processes have lower cross sections but with increasing LHC luminosity, it should in the near future become possible to extract the $W$ polarisation fractions.
Similarly, we expect the measurement of the $Z$ polarisation to be feasible at the LHC and have presented 
the relevant results.

\end{document}